\documentclass[12pt]{article}

\usepackage{graphicx, verbatim}
\usepackage{amsmath}
\usepackage{epsfig}
\usepackage{amsfonts}
\usepackage{amssymb}
\usepackage{cancel}
\usepackage{color}
\usepackage{float}
\restylefloat{table}

\newcommand{\be}{\begin{equation}}
\newcommand{\ee}{\end{equation}}
\newcommand{\bea}{\begin{eqnarray}}
\newcommand{\eea}{\end{eqnarray}}

\newcommand{\nn}{\nonumber}

\newcommand{\lp}{\left(}
\newcommand{\rp}{\right)}
\newcommand{\drbar}{$\overline{\rm DR}$}
\begin{document}
 
\begin{flushright}
DESY 15-247\\
\end{flushright}

\vskip 8pt

\begin{center} {\bf \LARGE {Detectable Gravitational Waves from \\
      Very Strong Phase Transitions \\ in the General NMSSM\\}}
\end{center}

\vskip 12pt

\begin{center}
{\bf Stephan J.~Huber$^a$, Thomas Konstandin$^b$, Germano Nardini$^{b,c}$, \\Ingo Rues$^b$ }\\[3mm]
{$^a$\em Department of Physics and Astronomy, University of Sussex, Falmer, Brighton BN1 9QH, U.K.}\\ {$^b$\em DESY, Notkestr.~85, D-22607 Hamburg, Germany } \\
{$^c$\em Albert Einstein Center for Fundamental Physics, Institute for Theoretical Physics, University of Bern, Sidlerstr.~5, CH-3012 Bern, Switzerland}\\
\end{center}

\vskip 20pt

\begin{abstract}
\vskip 3pt

\noindent

We study the general NMSSM with an emphasis on the parameter
regions with a very strong first-order electroweak phase transition
(EWPT).  In the presence of heavy fields coupled to the Higgs
sector, the analysis can be problematic due to the existence of sizable
radiative corrections. In this paper we propose a subtraction scheme
that helps to circumvent this problem. For simplicity we focus on a
parameter region that is by construction hidden from the current
collider searches. The analysis proves that (at least) in the
identified parameter region the EWPT can be very strong and striking
gravitational wave signals can be produced. The corresponding
gravitational stochastic background can potentially be detected at the
planned space-based gravitational wave observatory eLISA, depending on
the specific experiment design that will be approved.

\end{abstract}

\newpage

\section{Introduction\label{sec:intro}}

First-order phase transitions can establish testable links between
cosmology and particle physics. This is particularly interesting for
the electroweak phase transition (EWPT) that relates to the properties
of the Higgs sector which is currently tested at the LHC. Well known
examples for this link are electroweak 
baryogenesis~\cite{Kuzmin:1985mm} and gravitational wave (GW)
production~\cite{witten}.

In the Standard Model the EWPT is not of first order~\cite{nonpert}
but this behaviour can change in extensions of the standard model (SM). For
instance, a strong EWPT is possible in ultraviolet (UV) embeddings
that delay the SM-like electroweak symmetry breaking to
temperatures below $\mathcal O($10\,GeV)~\cite{Randall:2006py}. In
most of the cases, however, the EWPT turns to be strong because of new
electroweak-scale fields that extend the Higgs sector or induce radiative
modifications to it.

Among the plausible UV completions of the SM, supersymmetric theories
-- which have both an extended Higgs sector and new fields coupled to
it -- are candidates, that, in principle, can naturally lead to a
strong EWPT. In the Minimal Supersymmetric SM (MSSM), detailed studies
of the EWPT have been carried out~\cite{mssm1, mssm1bis}. It turns out that the
measurement of the Higgs mass, on top of LHC constraints on
new physics, is cornering the scenario into a parameter region that is in
tension with naturalness and collider data~\cite{mssm2}. It is hence worth
studying the electroweak symmetry breaking dynamics in supersymmetric
extensions where the 125-GeV Higgs mass can be accommodated more
naturally.

When looking for supersymmetric theories (and their parameter
regions) that possibly have a strong EWPT, it is a good guiding principle 
to start from non-supersymmetric extensions of the SM with a strong EWPT.  
These non-supersymmetric theories may be viewed as low energy approximations 
to the supersymmetric ones. The scenario we have in mind is a supersymmetric 
extension of the SM with most superpartners in the TeV region.  The difficulty lies in
rephrasing the successful low-energy (electroweak scale) parameter regions in terms of
the high-energy (TeV-scale) ones. In practice, one has to ensure that the successful low-energy
parameter configurations satisfy the supersymmetric
constraints at the high scale. These relations 
can be strongly modified by large radiative corrections induced by heavy
superpartners.

A strategy to deal with these two issues is to analyze the EWPT after
employing a ``match and run" procedure (cf.~e.g.~\cite{mssm1bis,
  Carena:2008rt}). The disadvantage of this technique is that it is
precise only when the heavy fields are much above the TeV scale. Since
present LHC bounds on new physics do not yet require such large
masses~\cite{LHCsearchesSUSY, LHCsearchesEXO}, in this paper we
propose an approach useful to determine the EWPT in scenarios where
the radiative corrections of the heavy fields are sizable but not so
large as to require a renormalization group (RG) resummation. This method
is based on a renormalization prescription that shares some
similarities with the on-shell scheme. In particular, we apply this
method to determine the EWPT and the corresponding GW
production in an illustrative supersymmetric theory with heavy
particles. The choice of this theory is motivated by 
the simple non-supersymmetric SM extensions exhibiting a {\it very} strong EWPT.

In many of the SM extensions with a strong EWPT, the barrier that is
required between the electroweak symmetric and broken minima of the Higgs
potential, is produced by thermal effects or quantum corrections. In
these cases, the EWPT typically exhibits small supercooling and latent
heat, i.e.~is not very strong. The contrary tends to occur in models
where the electroweak breaking minimum is close to metastability with the
unbroken phase, and the height and width of the barrier between the electroweak
symmetric and broken phases is generated by tree-level terms. These terms
in fact can easily induce a large jump of the order parameter during the transition and
lead to a regime of large supercooling.

The most simplistic model along these lines is the SM extended by a real scalar singlet.
In its minimal form with a $Z_2$
symmetry, the tree-level scalar potential of the Higgs and singlet
fields, $h$ and $s$, reads
\be
V(h, s) = - \frac12 \mu^2 h^2 - \frac12 \mu_s^2 s^2 
+ \frac{\lambda}{4} h^4 + \frac{\lambda_s}{4} s^4 + \lambda_m s^2 h^2 \, .
\ee
For a suitable range of parameters, this potential can be written as
\be
\label{eq:pot_Z2_singlet}
V(h, s) = \frac{\lambda}{4} 
\lp h^2 + \alpha^2 s^2 - v_h^2 \rp^2
+ \bar\lambda_m s^2 h^2 +  \frac12 \bar \mu_s^2 s^2 \, .
\ee
For $\bar\lambda_m$ and $\bar \mu_s^2$ small, the potential has an electroweak
broken minimum at $\langle \{h,s\}\rangle_{\cancel{EW}} =\{ v_h,0\}$
as well as electroweak symmetric minima at $\langle \{h,s\}\rangle_{EW}=\{0,
\pm \bar v_s\}$ with $\bar v_s\simeq v_h/\alpha$. The parameter $\bar
\mu_s^2$ controls the potential difference between the broken and
symmetric phase while the parameter $\bar\lambda_m$ controls the
height of the potential barrier between the two phases. For a suitable
set of parameters, the model exhibits a very strong phase transition
while particle phenomenology is in accord with all collider
constraints~\cite{LHCsingl, LHCsingl2}.

In fact, for some parameter values the potential
(\ref{eq:pot_Z2_singlet}) predicts a two-stage phase transition: At
very large temperatures the ground state of the system breaks neither
the electroweak symmetry nor the $Z_2$ symmetry. At
temperatures around the electroweak scale, the $Z_2$ symmetry is first
broken by a vacuum expectation value of the singlet field ($\langle h
\rangle_{EW} = 0,\, \langle s \rangle_{EW} \not= 0$).  At even lower
temperatures the electroweak symmetry is broken whereas the $Z_2$ symmetry is
restored ($\langle h \rangle_{\cancel{EW}} \not = 0,\, \langle s \rangle_{\cancel{EW}} =0$). It is
this second phase transition that is very strong and can lead to
cosmological implications.

Motivated by this feature of the singlet extension of the SM, in the
present work we study the EWPT in one of its possible UV embeddings:
the general next-to-minimal supersymmetric extension of the SM
(NMSSM)~\cite{Barger:2006dh}.  The aim is to identify the parameter
region with a very strong EWPT along the lines of the potential
(\ref{eq:pot_Z2_singlet})~\footnote{By a {\it very strong} phase
  transition we mean a phase transition that requires a large
  supercooling and hence generates GWs with amplitudes detectable at
  experiments such as eLISA~\cite{eLISAreport}. In sensible
  supersymmetric models a very strong EWPT was found only before the
  LHC bounds~\cite{NMSSMpreLHC+GW, oldNMSSM}.  Further findings on
  strong (but not very strong) EWPTs in singlet extensions of the MSSM
  have been reported in refs.~\cite{NMSSMpreLHC, Kozaczuk:2014kva}.}.
That this is possible is not guaranteed, since, as stated above,
supersymmetry implies constraints between different couplings of the
model, and predicts new particles whose phenomenology may be in
conflict with present experimental limits. To accommodate these
limits, some fields need to be heavy and their radiative corrections
must be kept under control to avoid the destabilization of the
tree-level results. We circumvent this issue by employing the
subtracting-scheme approach mentioned above.

The paper is organized as follows. In section~\ref{sec:model} we
introduce the general NMSSM and we identify a parameter region that is
promising for a two-step EWPT. Since this region involves some heavy
particles (around the TeV scale), in section \ref{sec:toy} we review how to deal with the
sizable radiative corrections induced by these heavy fields. 
In section \ref{sec:1loop} we describe an approach
useful to study the EWPT in the presence of heavy
fields. Section~\ref{sec:spectrum} and section~\ref{sec:pheno} discuss
the particle phenomenology of the parameter region identified in
section~\ref{sec:model}, and provide some parameter configurations
that are safe from plausible forthcoming LHC
constraints. Section~\ref{sec:num} contains the analysis of the EWPT
performed for some benchmark points, whereas section \ref{sec:GWs}
explains the corresponding gravitational backgrounds and their
detection perspectives at the forthcoming  space-based gravitational wave observatory eLISA.
Section~\ref{sec:dis} is devoted to some final remarks and
conclusions.  

\section{Tree-level analysis of the phase transition\label{sec:model}}

A supersymmetric theory that potentially reproduces the singlet
extension of the SM at low energy, is the general
NMSSM~\cite{Barger:2006dh}. In this supersymmetric theory all
renormalizable interactions respecting gauge symmetry and $R$ parity
are allowed. The superpotential involving the superfields of the
singlet and the Higgs doublets, $\hat S$, $\hat H_u$, and $\hat H_d$,
is
\be
W = L_1 \hat S 
+ \mu \hat H_u \hat H_d + \frac12 M_S \hat S^2
+ \lambda \hat H_u \hat H_d \hat S
+ \frac13 \kappa  \hat S^3
+ \cdots ~,
\ee
where terms with quark and lepton superfields have been omitted.

Including all soft terms, the potential of $h_u=\operatorname{Re}
[H_u^0/\sqrt{2}]$, $h_d={\rm Re} [H_d^0/\sqrt{2}]$ and $s={\rm Re}
[S/\sqrt{2}]$ reads
\bea
\label{eq:pot_init}
V_0 &=& \frac12 m_{H_d}^2 h_d^2 + \frac12 m_{H_u}^2 h_u^2 + \frac12 (B_S + m_S^2) s^2  \nn \\
&& + \frac{1}{3\sqrt{2}} T_\kappa s^3 - B_\mu h_d h_u - \frac{1}{\sqrt2} T_\lambda h_d h_u s \nn \\
&& + \frac{1}{32} (g_1^2 + g_2^2) (h_d^2 - h_u^2)^2 + \frac{2}{\sqrt{2}} \xi_1 \, s\nn \\ 
&& + \lp L_1 + \frac{1}{\sqrt{2}}M_S s + \frac{\kappa}{2} s^2  - \frac{\lambda}{2} h_d h_u \rp^2 \nn \\
&&   + \frac12(h_d^2 + h_u^2) \lp \frac{1}{\sqrt{2}} \lambda s  +  \mu\rp^2 ~.
\eea
We assume all parameters to be real as we are not interested in
effects of CP violation.  Our notation follows the definitions employed
in the public code SARAH~\cite{Staub:2013tta} which we use to check some of our
results.

To reproduce at low energy the singlet $Z_2$-symmetric extension of
the SM, we push the parameters towards the regime where the potential
\eqref{eq:pot_init} resembles \eqref{eq:pot_Z2_singlet} once the heavy
fields have been decoupled. We hence impose the mass of the heaviest
CP-even eigenstate to be above the electroweak scale, and the singlet VEV in
the electroweak broken phase to be vanishing, $\langle s
\rangle_{\cancel{EW}}=0$.  The latter is generically achieved by a
shift-redefinition of $S$, which helps to identify the relevant terms
breaking the $Z_2$ symmetry.

At this stage, it is useful to fix some of the parameters of the
potential by specifying the electroweak breaking minimum.  We fix the
quantities
\be
\{ m_{H_d}^2, m_{H_u}^2, T_\lambda\} 
\ee
by rephrasing the electroweak broken phase as
\be
\label{eq:brokenPhase}
\langle \{ h_d, h_u, s\}\rangle_{\cancel{EW}} = 
\{ v_h \, \cos \beta, v_h \, \sin \beta, 0 \} ~,
\ee
with $v_h=246$\,GeV. This implies
\bea
m_{H_u}^2 &=& \tilde B_\mu \cot\beta - \mu^2 - (\lambda^2 v_h^2/2)\cos^2 \beta + (m_Z^2/2) \cos 2\beta ~,\\
m_{H_d}^2 &=& \tilde B_\mu \tan\beta - \mu^2 - (\lambda^2 v_h^2/2)\sin^2 \beta - (m_Z^2/2) \cos 2\beta ~,\\
T_\lambda &=& -\lambda M_s + 2 \lambda \mu /\sin 2\beta  ~,
\eea
where $\tilde B_\mu = B_\mu+\lambda L_1$.

With this set of parameters it is more transparent how to reproduce
the regime where the heaviest $CP$-even and $CP$-odd Higgses decouple.
This is achieved by sending $\tilde B_\mu \to \infty$ while keeping
$\tan\beta$ constant. In the original parameters, for fixed $v_h$ this
limit implies that also $m_{H_d}^2, m_{H_u}^2 \to \infty$.

In general, one cannot impose a $Z_2$ symmetry on the potential
(\ref{eq:pot_init}) with the electroweak minimum \eqref{eq:brokenPhase}. This
would imply $\mu = 0$ which is in conflict with realistic chargino
masses (cf.~section~\ref{sec:pheno})~\footnote{In this sense, our
  analysis does not apply straightforwardly to singlet extensions of
  the MSSM where the $\mu$ term is forbidden.}. However, if the $Z_2$
symmetry is only imposed in the electroweak symmetric phase (i.e.~$\langle h_u
\rangle_{EW} = \langle h_d\rangle_{EW} = 0$), one obtains the constraints
\be
T_\kappa = -3 M_S  \, \kappa \, , \quad
\xi_1 = -   L_1 M_S \, .
\ee
Interestingly, away from the symmetric minimum, the resulting scalar potential is $Z_2$ symmetric up to a term of the form
\be
\label{eq:Z2_breaking}
\frac{\lambda \, \mu}{\sqrt{2} \cos \beta \sin \beta}
\lp \sin \beta \, \cos \beta \, (h_u^2 + h_d^2) - h_d h_u\rp s \, .   
\ee
In the limit $\tilde B_\mu \to \infty$, the light linear combination of $h_u$
and $h_d$ corresponds to the excitation along $(h_u, h_d) \propto
(\sin \beta, \cos\beta)$, while the orthogonal combination is heavy
and its VEV goes to zero. Correspondingly, at low energies the term in
(\ref{eq:Z2_breaking}) vanishes and the effective scalar potential for
the light degrees of freedom displays an (approximate) $Z_2$ symmetry.

The potential in the symmetric phase is then characterized by the
coefficients of the quartic and quadratic terms only. In order to keep
as many physical parameters as possible fixed, it
is also advisable to set the singlet VEV in the electroweak symmetric phase. We
adjust $m_S$ by parametrizing the electroweak symmetric minima as 
\be
\label{eq:symmPhase}
\langle \{ h_u, h_d, s\} \rangle_{EW}
= \{ 0, 0, \pm \bar v_s \} \, ,  
\ee
which imposes
\be
m^2_S=-B_S-M^2_S-2\kappa L_1 - \kappa^2 \bar v_s^2 ~ .
\ee
Using this parametrization, the potential difference between the
broken and symmetric phases can be written as
\bea
\Delta V_0 &=& V_0(v_u,v_d,0)- V_0(0,0,\bar v_s) \nn \\
&=&\frac{1}{32} \lp \bar v_s^4 \kappa^2 
- 2 v_h^4 \lambda^2 \sin^2 2 \beta
-  (g_1^2 + g_2^2) v_h^4 \cos^2 2 \beta 
\rp \, .
\label{eq:pot_diff}
\eea
Hence $\bar v^2_s \kappa^2$ should not be too large in order to ensure
that the electroweak broken minimum is the global minimum of the
potential. This will be less constraining when
considering the one-loop potential, since the loop corrections seem to
lower the broken phase (and thereby lift the Higgs mass).  At the same
time, adjusting the parameter $\bar v^2_s \kappa^2$ allows to tune the
model to be close to metastability.

Of course, above $Z_2$ symmetry is not exact. In particular, the heavy
scalar and pseudoscalar Higgses still break this symmetry
explicitly. This means that quantum corrections will not respect the
$Z_2$ symmetry and neither thermal corrections will (even though the thermal corrections of the heavy particles are in general rather small).  To preserve the
$Z_2$ symmetry order by order in perturbation theory will hence
require a certain tuning in the parameters. 

In any case, since we are interested in the model in the limit of
heavy scalar Higgs, very different scales are involved in the
calculations.  One way to deal with this problem is to use dimensional
reduction (\drbar) regularization, to integrate out the heavy degrees of
freedom and then run the parameters down to the electroweak scale using
the RG equations of the effective theory. The disadvantage of this method
is that it is precise only when the heavy fields are well above the
TeV scale (unless also higher-dimensional operators are taken into
account, which makes the analysis very cumbersome).
Since in our case we do not deal with overly large logarithms, we here
propose a different approach: we adopt a regularization scheme where
the EWPT quantities are rather insensitive to radiative
corrections. Our specific choice will be discussed after briefly
reviewing the perturbative problem caused by heavy fields
in minimally subtracted renormalization schemes.

\section{Heavy fields in a toy model\label{sec:toy}}

As previously explained, it is important to have a reliable strategy
to study the EWPT in scenarios with heavy fields. These fields can in
fact induce large radiative corrections that completely spoil the
tree-level results, or even generate perturbative
problems~\footnote{Of course, the modification of the tree-level
  results is not a conceptual problem. However this situation often
  requires analyses based on scans over a wide range of parameters and
  the possibility of missing (tuned) satisfactory regions.}.  The main
problem compared to the usual analysis used in collider phenomenology
is that we require a scheme in which the perturbative expansion is
under control not only close to the broken phase but also in the
symmetric phase. Both regions are equally important in the phase
transition analysis.  In this section we discuss some ideas on how to
deal with this issue.

For simplicity we focus on a toy model with one heavy and one light
real scalar field. The Lagrangian of the model is
\bea
{\cal L} &=& \frac12 (\partial_\mu \phi)^2 + \frac12 (\partial_\mu \Phi)^2  \nn \\
&& - \frac12 m^2 \phi^2 - \frac12 M^2 \Phi^2 \nn \\
&& - \frac14 \lambda_\phi \phi^4 - \frac14 \lambda_\Phi \Phi^4 
- \frac12 \lambda \phi^2 \Phi^2 \, ,
\label{eq:toyL}
\eea
where the parameters are renormalized quantities and counterterms are omitted. The field
$\phi$ is light, whereas the field $\Phi$ is heavy, $m^2 \ll
M^2$. Both $m^2$ and $M^2$ are assumed positive such that the fields
do not develop any VEV.

\begin{figure}
\centering
\includegraphics[width=0.7\textwidth]{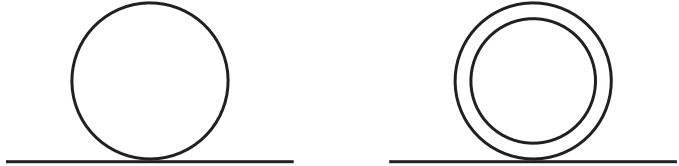}
\caption{The relevant self-energy diagrams of the soft field. The
  double line denotes the heavy field.}
\label{fig:one-loop-self}
\end{figure}

It is educative to analyze how the renormalized mass parameter $m^2$
is related to the mass observable $m^2_{\rm obs}$ in different
subtraction schemes.  We consider the one-loop correction to the
propagator of the light field $\phi$. Since the one-loop energy in the
current toy model is momentum independent, this only amounts to
studying the mass shift from the diagrams depicted in
figure~\ref{fig:one-loop-self}.

Using \drbar, the naive
one-loop relation between the renormalized and the observed mass
parameter reads
\bea
m^2_{\rm obs} = m^2  
 &+& \frac{3 \lambda_\phi}{16 \pi^2} 
m^2 \left[ \log(m^2/Q^2) - 1 \right] \nn \\
&& + \frac{\lambda}{16 \pi^2} 
 M^2 \left[ \log(M^2/Q^2) - 1 \right] \, .
\label{eq:one-loop-self}
\eea
Notably, this relation is problematic when $M^2$ is large. In
particular, for very large values of $M^2 \gg m^2_{obs} \sim Q^2$ the
equation has no solution for $m^2$ at all. This problem can be avoided when $Q$ is chosen 
such that the loop corrections are small. However,
such a strong dependence on the choice of $Q$ is undesirable and
finding such a peculiar $Q$ is a problematic task if there are several
light and/or heavy fields in the theory.

Notice also that this problem is not strictly related to large
logarithms. The problem already arises when the loop corrections are
larger than the observed mass, which does not necessarily lead to
large logarithms in the sense of 
\be
\label{eq:large_log}
3 \lambda_\phi \log (M^2/m^2) / 16 \pi^2 \gg 1\, .
\ee
Only if this relation holds, the resummation using the RG evolution is
relevant.

Finally, notice that using the naive relation \eqref{eq:one-loop-self} in 
the ``matching and run'' approach leads to the same problems.  Indeed 
the matching condition between above toy model and the low-energy effective 
theory (where only the light-field interactions are present) reads 
\be
\label{eq:matching}
\bar m^2 + \frac{3 \lambda_\phi}{16 \pi^2} 
\bar m^2 \left[ \log(\bar m^2/Q^2) - 1 \right] = \textrm{RHS of (\ref{eq:one-loop-self})} \, ,
\ee
where the renormalization scale $Q$ is of order $M$ and the parameter
$\bar m$ is the light mass parameter in the effective theory. Since
the left-hand side of this matching condition has the size of $m_{\rm
  obs}$, the problems above discussed arise also in this case. 

One way to remove these issues is to reorganize perturbation theory by
writing the normalized mass parameter as
\be
\label{eq:reorganize}
m^2 = (m^2 + \Delta m^2) - \Delta m^2 \, .
\ee
The sum of the two terms in the brackets is interpreted as the tree-level mass that is used in the propagator of
$\phi$. The last term is interpreted as a one-loop counterterm that
is grouped with the true one-loop quantum corrections.

The resulting relation for the mass turns out to be
\bea
m^2_{\rm obs} &=& (m^2 + \Delta m^2) \nn \\
 && + 
\left\{ \frac{3 \lambda_\phi}{16 \pi^2} 
(m^2 + \Delta m^2) \left[ \log([m^2 + \Delta m^2]/Q^2) - 1 \right] \right. \nn \\
&& \left. + \frac{\lambda}{16 \pi^2} 
M^2 \left[ \log(M^2/Q^2) - 1 \right] - \Delta m^2  \right\} \, . 
\label{eq:one-loop-reorganized}
\eea
Hence for the choice
\be
\label{eq:def_m2phi}
\Delta m^2 = \Delta m_\Phi^2 \equiv 
\frac{\lambda}{16 \pi^2} 
M^2 \left[ \log(M^2/Q^2) - 1 \right] \, ,
\ee
one obtains
\bea
m^2_{\rm obs} &=& (m^2 + \Delta m_\Phi^2) \nn \\
 && + \frac{3 \lambda_\phi}{16 \pi^2} 
(m^2 + \Delta m_\Phi^2) \left[ \log([m^2 + \Delta m_\Phi^2]/Q^2) - 1 \right] \, ,
\label{eq:one-loop-resum}
\eea
and therefore
\be
m_{\rm obs}^2 = m^2 + \Delta m_\Phi^2\, + \textrm{small one-loop correction} \, .
\ee
This relation has a well-defined solution and the hierarchy problem is
explicit. It is also obvious from this expression that as soon as
$\Delta m_\Phi^2$ approaches $m^2$ an expansion in $\Delta m_\Phi^2\ll
m_\Phi^2$ (which would mimic \eqref{eq:one-loop-self}) is not allowed.

This reorganization of perturbation theory is very close in spirit to
on-shell renormalization. There, one chooses renormalization
conditions and then ensures order by order in perturbation theory
using counterterms that the renormalization conditions are still
fulfilled. The difference to the reorganization of perturbation theory
using (\ref{eq:reorganize}) is that the tree-level mass parameter is
still $m^2$ rather than $m^2 + \Delta m^2$. Even though in the current
example both procedures are equivalent, the reorganization of
perturbation theory in (\ref{eq:reorganize}) has advantages. For
example, resummation at finite temperature has to be done using
this reorganization in order to avoid temperature dependent bare
mass parameters.

\vskip 0.5 cm

\begin{figure}
\centering
\includegraphics[width=0.5\textwidth]{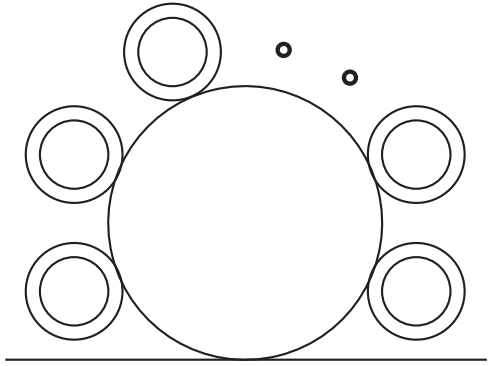}
\caption{Resummation of Daisy diagrams.}
\label{fig:daisies}
\end{figure}
That reorganizing perturbation theory resolves the issues indicates
that the problem actually stems from the breakdown of perturbation
theory. The class of diagrams that have to be resummed is depicted in
figure~\ref{fig:daisies}. The corresponding contribution is proportional
to
\bea
&& \hskip -2 cm
\int d^dp \, \frac{1}{p^2 - m^2} \sum_n \left[ \frac{\Delta m_\Phi^2}{p^2 - m^2}\right]^n \nn \\
&=& \int d^dp \, \frac{1}{p^2 - m^2 - \Delta m_\Phi^2} \nn \\
&=& \frac{1}{16 \pi^2} (m^2 + \Delta m_\Phi^2)
\left[ \log ([m^2 + \Delta m_\Phi^2]/Q^2) - 1\right] \, ,
\eea
where we used again the definition $\Delta m_\Phi^2$ for the loop contribution of the heavy field
as in (\ref{eq:def_m2phi}). Accordingly, the relation (\ref{eq:one-loop-self}) turns again into
\bea
m^2_{\rm obs} &=& (m^2 + \Delta m_\Phi^2) \nn \\
 && + \frac{3 \lambda_\phi}{16 \pi^2} 
(m^2 + \Delta m_\Phi^2) \left[ \log([m^2 + \Delta m_\Phi^2]/Q^2) - 1 \right] \, .
\label{eq:one-loop-resum-DRbar}
\eea
Similarly, using the resummed one-loop expression for the light mass,
the matching condition (\ref{eq:matching}) in the effective theory
becomes almost trivial~\footnote{A detailed calculation of the
    matching condition of the mass parameters in the presence of
    several heavy fields can be found in ref.~\cite{Masina:2015ixa}
    where the MSSM case is explicitly described.}
\be
\bar m^2 = m^2 + \Delta m_\Phi^2\, .
\ee
Notice that reorganizing perturbation theory is equivalent to this
resummation.  In fact, using counterterms does not only correspond to
the Daisy resummation on the on-loop level. The counterterm in the 
reorganized theory will remove all the Daisy diagrams we have resummed 
in the last paragraph.

\begin{figure}
\centering
\includegraphics[width=0.5\textwidth]{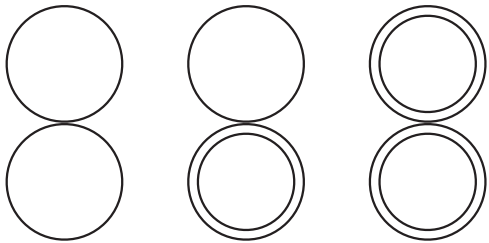}
\caption{The relevant two-particle-irreducible vacuum diagrams. The propagators are full propagators.}
\label{fig:2pi}
\end{figure}
Resumming relevant (hand-picked) contributions is only one option. After all, it
is not guaranteed that the resummed diagrams really contain all contributions that 
are relevant to improve the convergence of the perturbative series. A more sophisticated approach 
would be to use the two-particle-irreducible effective action~\cite{Cornwall:1974vz}. The corresponding diagrams are depicted in figure~\ref{fig:2pi}. This time the propagators are understood to be the full propagators. In the current toy model, the only relevant diagram for resummation is the mixed light/heavy diagram while the other two diagrams only contribute small shifts 
in the masses. Only considering the mixed diagram then leads to the same result as the explicit resummation above. 

\vskip 0.5 cm

In summary, only removing divergences and adjusting the mass
parameters will not lead to reliable results once the loop contributions
to the self-energies approach the order of the physical masses. In
this regime, the convergence of the perturbative series is much better
in renormalization schemes that use counterterms and are closer to
on-shell renormalization. Alternatively, the Daisy diagrams of
perturbation theory can be resummed. Since it is easier to implement,
we will use the former approach in the following.

\vskip 0.5 cm

A similar discussion applies also to the cases where the light
  scalar particle obtains a VEV. In these scenarios, the VEV is
often used as renormalized (input) parameter instead of the squared
mass parameter.  The VEV, which is obtained by solving the tadpole
  equation, is then kept fixed order by order in the perturbative
expansion. The loop corrections of this equation contain the mass of
the light scalar in terms of its mass parameter and its VEV.  Also the
heavy-field masses expressed in terms of the VEV, enter in the loops
and they induce large contributions to the tadpole equation.  To
compensate these contributions, the mass parameter has to be
drastically modified, with the consequent dangerous effect in the loop
involving the light field. On the contrary, if the VEV is kept
  fixed by introducing a counterterm that cancels the large radiative
  contributions to the tadpole equation, the issue is mitigated.  In
particular, such a counterterm amounts to the contribution that
  keeps the mass parameter fixed (cf.~eq.~\eqref{eq:def_m2phi}) up to
  tiny radiative corrections as e.g.~the one to $\lambda_\phi$ (which
  is small when eq.~\eqref{eq:large_log} holds). In this regime,
  therefore, applying counterterms to fix the VEV in the tadpole
  equations is perturbatively as good as the resummation above. This
  is the scheme usually employed in NMSSM
  phenomenology~\cite{Degrassi:2009yq, Staub:2010ty, Ender:2011qh,
    Graf:2012hh, Staub:2015aea}.
 
As mentioned at the beginning of the section, when studying the 
phase transition, it is not enough to ensure the convergence of the 
perturbation theory in the broken phase. The symmetric phase is equally 
important and one needs a subtraction scheme in which observables (in particular the VEVs and  masses) are stable against quantum corrections also in the symmetric phase. 
Using counterterms for the three tadpole equations only in the broken phase 
will not achieve this
and lead to sizable corrections to the effective potential in the symmetric 
phase at one-loop level. Our strategy will be to introduce counterterms for all 
soft supersymmetry breaking parameters. These will be fixed by conditions on the 
effective potential that also involves the symmetric phase. 
In particular, we would like to preserve (at
least approximately) the $Z_2$ symmetry on loop level. The precise 
renormalization conditions are discussed in the next section.

\section{One-loop construction\label{sec:1loop}}

In the one-loop analysis we use the scalar potential
(\ref{eq:pot_init}), supplemented by the one-loop effective potential
obtained using \drbar~scheme,
\be
V_{CW}(h_u,h_d,s) = \sum_{\textrm{species}}
\pm \frac{n_i}{64 \pi^2} 
m_i^4 \left[ \log(m_i^2/Q^2) - \frac32 \right] \, ,
\label{eq:CW_pot}
\ee
where $n_i$ is the number of degrees of freedom of each species (with
a negative sign for fermions) whose squared mass $m_i^2$ is expressed as
a function of the $h_u,h_d,s$ backgrounds fields. Besides, we need 
to add the effect of the different counterterms
\bea
V_{\textrm{cnt}} &=& \frac12 \Delta m_{H_d}^2 h_d^2 + \frac12 \Delta m_{H_u}^2 h_u^2 
+ \frac12 \Delta m_S^2 s^2 \nn \\ 
&& + \frac{1}{3\sqrt{2}} \Delta  T_\kappa s^3  - \frac{1}{\sqrt2} \Delta  T_\lambda h_d h_u s \nn \\
&& - \Delta B_\mu h_d h_u  + \frac{1}{\sqrt{2}} \Delta \xi_1 \, s  \, .
\eea
These counterterms are considered to be of one-loop order and hence do
not change the masses used in the one-loop contributions
(\ref{eq:CW_pot}). Moreover, they are in accordance with softly broken supersymmetry.

Since we want to study the phase transition of the system, we choose
renormalization conditions that do not only hold the VEVs in the
broken phase fixed but also the VEVs in the symmetric phase.  The
counterterms are determined imposing on the full one-loop potential
\be
V_{\textrm{1-loop}} = V_0 + V_{CW} + V_{\textrm{cnt}} \, ,
\ee
the following tadpole conditions in the electroweak broken phase (\ref{eq:brokenPhase})
\be
\left. \partial_{h_u} V_{\textrm{1-loop}} \right|_{\cancel{EW}} = 0 \, , \quad
\left. \partial_{h_d} V_{\textrm{1-loop}}\right|_{\cancel{EW}} = 0 \, , \quad
\left. \partial_{s} V_{\textrm{1-loop}}\right|_{\cancel{EW}} = 0 \, , \quad
\ee
and the following tadpole condition in the two electroweak symmetric phases
(\ref{eq:symmPhase})
\be
\left. \partial_{s} V_{\textrm{1-loop}}\right|_{EW}  = 0 \, . \quad
\ee
Furthermore, we impose in the symmetric phase
\be
V_{\textrm{1-loop}} \left|_{s = \bar v_s}
 = V_{\textrm{1-loop}} \right|_{s = - \bar v_s} \, .
\ee
This leaves us with one free counterterm that we could in principle use 
to minimize the mixing between the light Higgs particle and the scalar singlet 
in the broken phase. In practice this mixing is small due to the $Z_2$
symmetry on tree level and we use $\Delta B_\mu = 0$. 

At first sight, the above choice for the renormalization conditions
seems arbitrary and unphysical. However, it turns out that these
choices keep almost all masses relatively stable against loop
corrections (cf.~section~\ref{sec:toy}). This of course helps in
identifying the parameter region where the low-energy theory
resembles the singlet extension of the SM with a very strong EWPT.
 Besides, also the properties of the phase transition are
observable, and it is desirable that they are not heavily influenced
by loop corrections.  Moreover, the above construction reduces the
dependence on the renormalization scale $Q$ tremendously compared to
the plain (i.e.~not resummed) \drbar~scheme.

\section{Tree-level spectrum\label{sec:spectrum}}
\noindent
The main features of the spectrum can be captured at tree level after
imposing the minimization conditions explained in the previous
section. 

In the basis $\{h_1,h_2,h_3 \}$ with $h_1=s$, $h_2=h_{d}\cos\beta +
h_{u}\sin\beta$ and $h_3=h_{d}\sin\beta - h_{u}\cos\beta$, the
symmetric CP-even squared mass matrix is given by 
\bea
\label{eq:tree_scalars}
\mathcal{M}_{S,11}^2&=&-\kappa^2 \bar v_s^2 - \kappa\lambda v_h^2\sin\beta \cos\beta + \lambda v_h^2/2\nonumber~,\\
\mathcal{M}_{S,22}^2&=& m_Z^2\cos^2 2\beta + (\lambda^2 v_h^2/2)\sin^2 2\beta\nonumber~,\\
\mathcal{M}_{S,33}^2&=& m_{A_2}^2 + (2 m_Z^2-\lambda^2v_h^2)(1-\cos 4\beta)/4\nonumber~,\\
\mathcal{M}_{S,12}^2&=& 0 \nonumber~,\\
\mathcal{M}_{S,13}^2&=&\sqrt{2}\lambda \mu v_h \cot (2\beta) ~, \nonumber\\
\mathcal{M}_{S,23}^2&=& \sin (4\beta)(2 m_Z^2 - \lambda^2 v_h^2 )/4 ~,
\eea
with $m_{A_2}^2=\tilde B_\mu/(\sin\beta \cos\beta)$ and
$m_Z^2=(g_1^2+g_2^2) v_h^2/4$.

A similar procedure can be applied also to the pseudoscalar and
charged Higgs squared masses. After a rotation leading to
$A_2=h_{uI}\cos\beta + h_{dI} \sin\beta$ and $G^0=-h_{uI}\cos\beta
+h_{dI} \sin\beta$ (with $h_{uI}={\rm Im}[H_u^0]/\sqrt{2},
h_{dI}={\rm Im}[H_d^0]/\sqrt{2}$), and omitting the entries corresponding
to the Goldstone boson $G^0$, the CP-odd squared mass matrix in the
basis $\{A_1={\rm Im}[S]/\sqrt{2}, A_2\}$ turns out to be
\bea
\label{eq:tree_pseudoscalars}
\mathcal{M}_{P,11}^2&=& -\kappa^2 \bar v_s^2 + \kappa\lambda v_h^2 \sin\beta\cos\beta-2 B_S-4\kappa L_1 + \lambda^2 v_h^2/2 ~,\nonumber\\
\mathcal{M}_{P,12}^2&=&  \sqrt{2}\lambda v_h \lp \mu/\sin(2\beta)-M_S\rp \nonumber~,\\
\mathcal{M}_{P,22}^2&=& m_{A_2}^2 ~.
\eea
Performing the same rotation on the squared mass matrix of the charged
Higgses, one obtains a massless linear combination, which corresponds to
the charged Goldstones, and the physical charged Higgs, $H^\pm$, 
with mass 
\be 
M^2_\pm=m_{A_2}^2 + m_W^2- \lambda^2 v_h^2/2 ~.
\ee

Notice that in the decoupling limit $m_{A_2}\to \infty$, the
off-diagonal entries in $\mathcal M_S^2$ and $\mathcal M_P^2$ are negligible and
correspondingly the fields $h_1, h_2,h_3,A_1$ and $A_2$ are a good
approximation to the (tree-level) eigenstates fields whose masses are
the diagonal elements of the matrices. In this case a judicious choice
of some parameters, e.g.~$B_S$, establishes the (tree-level)
mass hierarchy between the singlet-like $CP$-even and $CP$-odd
Higgses. Moreover, due to the negligible mixing between the
singlet-like and SM-like Higgses, radiative corrections coming from
sfermions are not expected to change such a hierarchy. These corrections
are instead important for the SM-like Higgs, $h_2$. In fact they can
efficiently lift $m_{h_2}^2$ to the experimental value
$\sim125$\,GeV. We use this property to set the squark and slepton
masses, assuming vanishing trilinear terms and masses all degenerate
(although in the regime $\tan\beta\lesssim 10$ we will consider, only
stops are relevant for the Higgs boson mass and the EWPT). In practice this implies all sfermions are at the TeV scale.

In concrete cases, $m_{A_2}$ does not necessarily need to be much
above the electroweak scale to lead to the decoupling scenario characterized by
a low-energy $Z_2$ symmetry.  A reasonable hierarchy of $m_{A_2}^2$
versus $m_Z^2$ and $\lambda^2v_h^2$ is sufficient to guarantee a
SM-like Higgs $h_2$ without peculiar cancellations in $\mathcal
M_{S,23}^2$. Also the states $h_1$ and $h_3$ have a tiny mixing if
$\mu v_h$ is small enough compared to $m_{A_2}^2$. Finally, if $A_1$
is light, its mixing with $A_2$ is negligible for $M_S$ sufficiently
small (although $A_1$ is not required to be small to the aims of the
EWPT). In conclusion, $A_2$ (and $A_1$) can be just above the electroweak scale
without peculiar cancellations in the entries $\mathcal M_{S,23}^2$
and $\mathcal M_{S,13}^2$ (and $\mathcal M_{P,12}^2$) if the Higgsino
mass term $\mu$ is (and the singlino mass term $M_S$ are) small. This
choice would have implications for the neutralino and chargino mass
spectra, which can be deduced from the mass 
matrices
\begin{equation} 
\label{eq:mnmass0}
{\mathcal M}_{\widetilde\chi^0} = \left( 
\begin{array}{ccccc}
M_1 &0 &-\frac{1}{2} g_1 v_d  &\frac{1}{2} g_1 v_u  &0\\ 
\cdot &M_2 &\frac{1}{2} g_2 v_d  &-\frac{1}{2} g_2 v_u  &0\\ 
\cdot &\cdot  &0 &- \mu  &-\frac{1}{2}  v_u \lambda \\ 
\cdot  & \cdot & \cdot  &0 &-\frac{1}{2}  v_d \lambda \\ 
\cdot & \cdot & \cdot  & \cdot  & M_S \end{array} 
\right) \, , 
\end{equation} 
%
%
\begin{equation} 
\label{eq:mnmasspm}
{\mathcal M}_{\widetilde\chi^\pm} = \left( 
\begin{array}{cc}
M_2 & \frac{1}{2} g_2  v_u \\ 
\frac{1}{2} g_2  v_d  & \mu 
\end{array} 
\right)~, 
\end{equation}
the usual NMSSM neutralino and chargino bases are employed and the electroweak
breaking minimum of eq.~\eqref{eq:brokenPhase} is assumed. Since gauginos do not play any
relevant role in the EWPT, for simplicity we assume all of them to be
degenerate at 1\,TeV.

\section{Phenomenological constraints\label{sec:pheno}}

\noindent
In the previous sections we have identified a region where the
tree-level scalar Higgs sector of the general NMSSM has an approximate
$Z_2$ symmetry at low energy. In this section we analyze the main
experimental constraints on this parameter region. This will guide us
to select some benchmark points to prove that in some simple
supersymmetric models, the present and foreseeable future experimental
bounds do not forbid a very strong EWPT  with a sizable
stochastic GW background.

As previously mentioned, for concreteness we fix the stops and
gauginos at around 1\,TeV. This is in agreement with stop-gluino
searches~\cite{LHCsearchesSUSY} and implies enough radiative
corrections~\cite{TuningNMSSM} to make $m_{h_2}$ compatible with the SM-like Higgs mass
measurements~\cite{Aad:2015zhl}.

Further constraints on $h_2$ come from the measurements on the 125-GeV
Higgs signal strengths~\cite{Aad:2015gba}. In general the field $h_2$ may depart
from the SM-like behavior because of two reasons: existence of new
decay channels and presence of sizable mixing with $h_1$ and
$h_3$. Both issues can be avoided in the identified parameter
region. Indeed no dilution of the $h_2$ branching ratios occurs if
$m_{h_1},m_{A_1},m_{\chi^0_1}>m_{h_2}/2$. Moreover, as discussed in
section~\ref{sec:spectrum}, $h_2$ almost does not mix with $h_3$
since $m_{A_2}$ is set to be in the decoupling limit~\footnote{In
  principle $h_2$ can be aligned to the SM-like Higgs even without
  decoupling $m_{A_2}$~\cite{Delgado:2013zfa}.}. On the other hand, at tree level
$h_2$ has no mixing with $h_1$ since $\mathcal M_{S,12}^2$ is
vanishing by construction. Radiative corrections generate
such a mixing~\footnote{We remind the reader that the low-energy $Z_2$ symmetry is not
  exact. Accordingly, $A_1$ or $h_1$ are no good dark matter
  candidates.}  but typically well within the 95\%~C.L.~experimental
limit, which requires $\sin^2\gamma<0.23$~\cite{Giardino:2013bma}, where $\gamma$ is defined
as
\be 
\tan 2\gamma=\frac{2\, \mathcal M^2_{S,12}}{\mathcal
    M^2_{S,22}-\mathcal M^2_{S,11}} ~.
\ee
As the numerical results of the
next section show, in the identified parameter region, $\tan 2\gamma$
turns out to be very small also at one-loop.

The tiny mixing of $h_1$ and $A_1$ with the Higgs doublets (as well as
the fact that the singlet does not acquire a VEV) is essential to
overcome also the experimental bounds on $h_1$ and $A_1$. The
literature has broadly investigated direct and indirect searches
sensitive to non-standard CP-odd and CP-even Higgs
bosons~\cite{LHCsingl2, Bomark, Tesi, Tania}. Besides the decays $h_2\to
A_1A_1 , h_1h_1$ which are kinematically closed in our case (note that
$h_3$ is heavy and thus rarely produced), the $h_1$ and $A_1$ decays
into SM particles are tightly constrained by the BaBar, LEP and LHC
analyses in a wide mass region of $m_{h_1}$ and
$m_{A_1}$. Nevertheless, none of these bounds seem to rule out
the identified parameter region with $m_{h_1},m_{A_1}\gtrsim 80\,$GeV
when $\mathcal M^2_{S,13}$ is kept reasonably below its critical value
corresponding to $\sin^2\gamma\simeq 0.23$~\cite{Tania}.
Finally, unless $m_{A_2}$ is extremely heavy, the requirement of not
too large Higgs-singlet mixing without tuning in the parameters, and the necessary boost of
  the tree-level mass $m_{h_2}$, implies a limit on $\mu$ and tends
to favor light Higgsinos close to the electroweak scale
(cf.~eqs.~\eqref{eq:tree_scalars} and
  \eqref{eq:tree_pseudoscalars}). For the same argument, also the
  singlino mass term $M_S$ should not be larger than $m_{A_2}$. This
  however allows for a hierarchy between the singlino and neutralino
  mass. In particular, if the singlino is sufficiently heavy, as well
as the gauginos, the Higgsino states are almost degenerate in
mass. Consequently the lighter chargino and the two lightest
neutralinos often exhibit multibody decays into pions, which are
challenging at the LHC but might be detected at the
  ILC~\cite{Berggren:2013vfa}. In this sense the benchmark points
with the Higgsino as lightest supersymmetric particle (LSP) are
  not problematic with respect to current bounds.

However, from a cosmological point view, the Higgsino as LSP is not
fully satisfactory since it cannot account for the observed DM relic
density. Among several further DM possibilities (e.g.~axion, axino or
gravitino), assuming a light singlino can solve the problem. For
instance for $M_S\sim \mu$ the LSP can yield the correct DM relic
  density while being safe from direct and indirect detection
  constraints~\cite{Enberg:2015qwa}.  The collider constraints on
this chargino-neutralino configuration are weak: due to the relatively
small mass splitting between $\chi_2^\pm$,$\chi_1^\pm$,$\chi_2^0$ and
$\chi_1^0$, charginos and neutralinos decay via off-shell gauge/Higgs
bosons whose products are soft and hence difficult to
trigger~\cite{Kim:2014noa}.  Notably, although we will not study
  numerically the case $M_S\sim \mu$, a priori it  seems
  compatible with the requirement of a very strong EWPT.

\section{Very strong EWPT\label{sec:num}}

In this section we present some numerical results on the strength
of the EWPT in the general NMSSM. We are not interested in providing a
comprehensive analysis of the ample parameter space but we content
ourselves with several benchmark points with a first-order EWPT. 

In order to avoid the phenomenological bounds discussed in
section~\ref{sec:pheno} we impose most of the supersymmetry breaking
parameters that are not relevant for the EWPT to be around the TeV
scale (in fact exactly 1 TeV if not noted otherwise).  This includes
gaugino masses, squark and slepton masses.  We also choose $B_\mu$ 
large to decouple some of the scalar degrees of freedom. For concreteness the parameter $M_S$ is assumed sizable,
which lifts the singlino mass well above the Higgsino mass with
  $\mu=300$ GeV (cf.~eq.~\eqref{eq:mnmass0}), whereas the squarks
  trilinear parameters and $L_1$ are set to zero for simplicity. The
  slepton and squark masses are assumed degenerate and adjusted to
  reproduce $m_{h_2}\approx 125\,$GeV at one loop. The parameters
  $m_S$ and $B_S$ establish the hierarchy between $h_1$ and $A_1$. 
Since the EWPT is not very sensitive to $B_S$ one can fix it to avoid any
  tension between $m_{A_1}$ and collider data.
Moreover, $m_{H_d}^2, m_{H_u}^2, m_S^2, T_\lambda, T_\kappa$
  and $\xi_1$ are fixed to enforce the approximate tree-level $Z_2$
  symmetry as described in section~\ref{sec:model}. 

Concerning the dimensionless parameters, we choose a combination of
  $\lambda$ and $\tan\beta$ that provides a relevant boost to the
  Higgs mass without spoiling the perturbativity of the theory below
  the grand unified scale~\cite{Ellwanger:2009dp}. Moreover, $\kappa$
  is set at a small value in order to have the electroweak phase as the global
  minimum of the potential even in the presence of large $\bar v_s$
  (cf.~eq.~\eqref{eq:pot_diff}).

With these choices, the only light particles in the
scalar sector (meaning masses much smaller than 1 TeV) are one
neutral Higgs, the singlet, one pseudoscalar, two neutralinos and one
chargino. The remaining particles, which are heavy, can lead to large one-loop
contributions to the potential of the light degrees of freedom. We
keep these radiative corrections under control by means of the
counterterms discussed in sections~\ref{sec:toy} and
\ref{sec:1loop}.

\begin{table}
\begin{center}
\begin{tabular}{ | c || c | c | c | c | }
  \hline                       
 & A & B & C & D \\
  \hline
  \hline
 $\bar v_s $ [GeV] 	& $307.5$	& $319.8$ 	& $323.5$ & $324.0$ \\
  \hline
\end{tabular}
\vskip 0.3cm
\begin{tabular}{ | c || c | c | c | c | }
  \hline                       
 & A - D \\
  \hline
  \hline
 tan $\beta$ 		& $5$ 		\\
 $\lambda$ 		& $0.7$ 	\\
 $\kappa$ 		& $0.015$	\\
 $L_1$ 			& $0$		\\
 $B_S$ [GeV$^2$] 			& $-250^2$ \\
 $\mu$ [GeV] 			& $300$	\\
  \hline
\end{tabular}
\caption{
\label{tab:params}
Parameters of the considered benchmark scenarios. All parameters but  $\bar v_s $ are
kept constant in the four benchmark points.}
\end{center}
\end{table}

\begin{table}
\begin{center}
\begin{tabular}{ | c || c | c | c | c | }
  \hline                       
tree & A - D\\
  \hline
  \hline
 $m_{h_1}$ 		& $93$ 	 \\
 $m_{h_2}$ 		& $96$ 	 \\
 $m_{A_1}$ 		& $373$ 	\\
 $m_{\chi_1^0}$ 	& $286$ 	\\
 $m_{\chi_2^0}$ 	& $310$  	\\
 $m_{\chi_1^\pm}$ 	& $296$ 	\\
  \hline 
\end{tabular}
\qquad\qquad
\begin{tabular}{ | c || c | c | c | c |  }
  \hline                       
1-loop & A - D \\
  \hline
  \hline
 $m_{h_1}$ 		& $91$ 	\\
 $m_{h_2}$ 		& $125.6$ 	\\
 $\sin^2 \gamma$ 	& $10^{-3}$ \\
\hline
\end{tabular}
\caption{
\label{tab:treeSpec}
Tree-level spectrum (left) and one-loop spectrum and mixing (right) of
the four benchmark scenarios quoted in table \ref{tab:params}. In the four scenarios
the light masses and the mixing angle $\gamma$ are the same within our
calculation uncertainties.}
\end{center}
\end{table}

\begin{table}
\begin{center}
\begin{tabular}{ | c || c | c | c | c | }
  \hline                       
      & A & B & C & D \\
  \hline
  \hline
 $T_{n}$ [GeV]	& $112.3$ 	& $94.7$	& $82.5$ & $76.4$ \\
 $\alpha$ 		& $0.037$ 	& $0.066$	& $0.105$ & $0.143$ \\
 $\beta/H$ 		& $277$ 	& $105.9$ 	& $33.2$ & $6.0$ \\
 $v_h(T_n)/T_{n}$        & $1.89$	& $2.40$	& $2.83$ & $3.12$\\ 	
\hline
\end{tabular}
\caption{
\label{tab:pt}
Characteristics of the EWPT in the benchmark points of table \ref{tab:params}.}
\end{center}
\end{table}

\begin{figure}
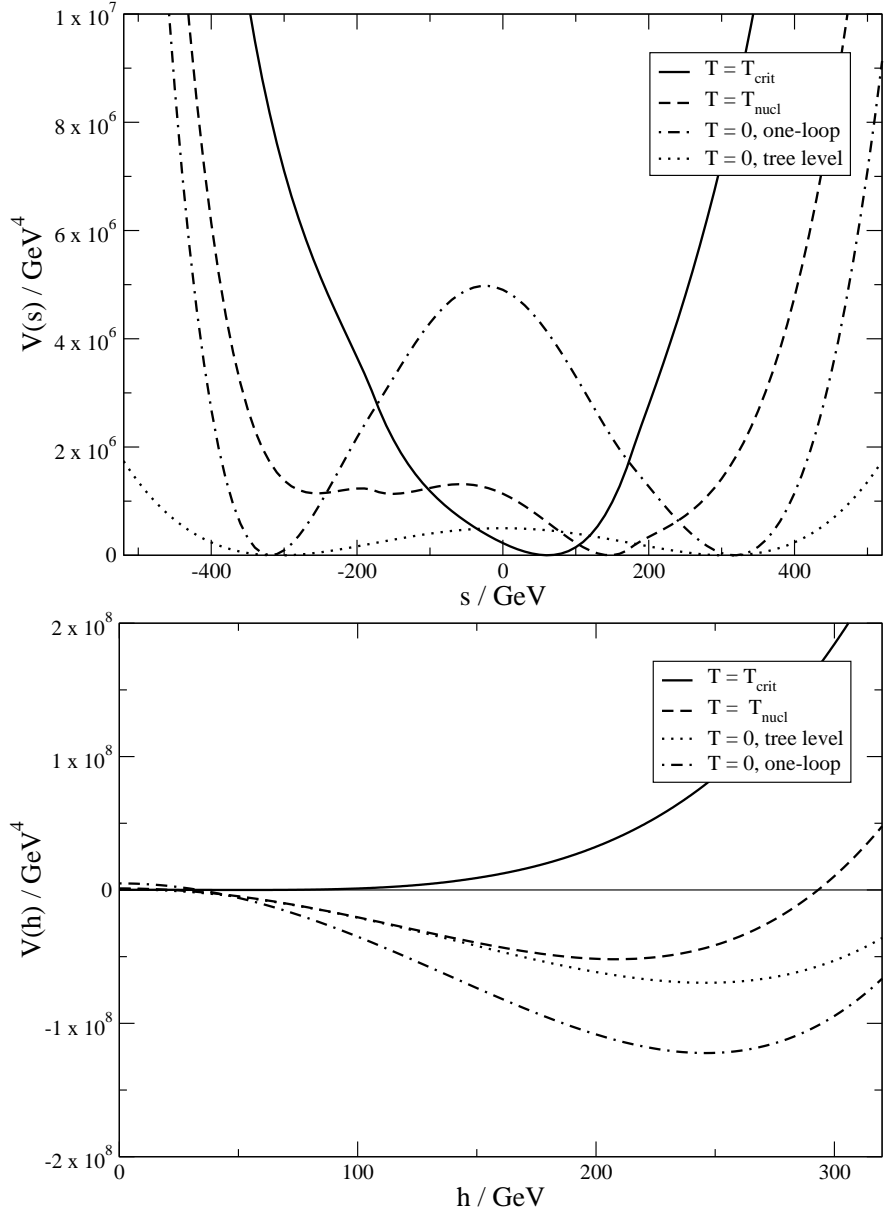

\centering
\includegraphics[width=0.85\textwidth, clip]{figs/pot0.eps}
\includegraphics[width=0.85\textwidth, clip]{figs/poth.eps}
\caption{The finite temperature potential in the benchmark point
  C. The upper [lower] plot shows the effective potential along the
  singlet direction $V(h=0,s)$ [along the SM-like Higgs direction
  $V(h,s=\left<S\right>_{\cancel{EW},T})$]. The potential minima at tree level and at one loop are
  the same as a consequence of the subtraction scheme we
  adopt.}
\label{fig:pot}
\end{figure}

In the following we analyze four benchmark points in parameter space
with phase transitions ranging from weakly first order to very strongly
first order and hence close to metastability. Table \ref{tab:params} displays the parameters we use in
each scenario, and table~\ref{tab:treeSpec} (left) shows the
corresponding tree-level masses of the light degrees of freedom.  Out
of these, the bosonic fields are potentially strongly affected by
the radiative corrections.  However, the Higgs and singlet masses
depend on the parameter combination $(B_S + m_S^2)$ while the pseudoscalar mass depends on $(B_S - m_S^2)$.  Accordingly, we use this
freedom to adjust the pseudoscalar mass freely and its precise
one-loop mass is therefore not very relevant for the EWPT.  The
scalar one-loop masses in turn are not free to chose and are given in
table~\ref{tab:treeSpec} (right). These are obtained (not on-shell
but at zero external momentum) from the one-loop effective potential.
It turns out that the 125-GeV Higgs has a tiny singlet component
($\sin^2 \gamma= 10^{-3})$, much below its experimental upper
bound, $\sin^2\gamma\lesssim 0.23$~\cite{Tania}. Also the spectrum is in accord with the collider constraints
discussed in the last section and is rather insensitive to changes in
$\bar v_s$.  This is due to the fact that the dependence 
of the scalar masses on $\bar v_s$ in (\ref{eq:tree_scalars}) and
(\ref{eq:tree_pseudoscalars}) is suppressed by factors of $\kappa$,
which is small as explained above.

The calculation of the tunneling action $S_3/T$ is in the current
model a multi-field problem. In order to simplify the analysis,
we assume a smooth path in field space that passes through the two minima and the
saddle point of the scalar potential. The nucleation temperature
$T_{n}$, the ratio of the vacuum energy to the thermal energy of the
plasma, $\alpha$, the quantity $\beta/H=\partial(S_3/T)/\partial \log
T$, and the strength of the EWPT $v_h(T_n)/T_n$  (where $v_h(T_n)$ is the VEV of the Higgs at the temperature $T_n$) are
then calculated by the usual one-dimensional over-/under-shooting
procedure~\cite{Coleman:1977py, Callan:1977pt, Linde:1980tt}. The
resulting tunneling action can be quite large due to the fact that
the scalar fields have to cover a relatively long distance in field
space.  Therefore the scalar potential only requires
a rather small potential barrier at the nucleation temperature (see
figure~\ref{fig:pot}).

As anticipated, the model displays a very strong first-order phase
transition. By increasing the singlet VEV in the symmetric phase, $\bar v_s$,
we dial the strength of the phase transition.  According to
(\ref{eq:pot_diff}), changing $\bar v_s$ has a direct impact on the
potential difference between the symmetric and broken phases. At a
certain value $\bar v_s \simeq 325$ GeV the system will not be able to
overcome the potential barrier and reside in the symmetric
phase. Close to metastability (scenario D), we observe large Higgs
bubbles ($\beta/H \lesssim 10$) and sizable latent heat ($\alpha
\gtrsim 0.1$).

In the next section, we discuss the expected GW signal of the 
benchmark scenarios.

\section{Gravitational waves\label{sec:GWs}}

In this section we describe the calculation of the GW production in
the EWPT of the benchmark points of table \ref{tab:params}. For this we have to take into account
that the phase transition does not occur in an empty universe, but rather in the
presence of a hot plasma of relativistic particles (`fluid').
The calculation will be
performed in two ways. 

Firstly, we will treat the expanding bubbles and the fluid they drag
with them as a system of vacuum bubbles only \cite{Huber:2008hg}. We
model the phase transition as collision of these `vacuum bubbles', and
compute the resulting gravitational wave signal.  In a second stage,
we model the fluid in a more detailed manner. The phase transition
then leads to the creation of sound waves, which in turn produce
gravitational waves \cite{Hindmarsh:2013xza,
  Hindmarsh:2015qta}. Finally the resulting gravitational waves
signals will be compared to the sensitivities of the experiment that
will be possibly launched in the forthcoming ESA Gravitational Wave
Mission, (preliminarily) called eLISA.

\subsection{Gravitational wave signal from bubble collisions}

The calculation of the production of GWs from bubble collisions will
follow~\cite{eLISAreport}. For this we only need three
characteristics of the phase transitions in each benchmark point:
$\alpha$, $\beta/H$ and $v_w$. The former two quantities are
calculated in the previous section and quoted in table \ref{tab:pt}. The last
parameter is the expansion velocity of the bubbles. The
dynamics of the bubble walls are quite complicated and have been
studied in \cite{MoAndPro, John:2000zq, Megevand:2009gh, energybudget, Megevand:2012rt, Huber:2013kj, Konstandin:2014zta, Kozaczuk:2015owa}. These
calculations have been also employed to determine $v_w$ in some NMSSM
scenarios where the EWPT is not very
strong~\cite{Kozaczuk:2014kva}. Fortunately there is a simple
criterion to see if the bubble wall approaches the speed of light: it
occurs when the effective potential in the mean-field
approximation at $T=T_n$ is larger in the unbroken phase than in the broken
phase~\cite{Bodeker:2009qy}. Since this condition is fulfilled for our
benchmark points, we can consider $v_w \simeq 1$. 

If the bubble walls can run away and reach very large gamma factors in the case of very strong phase transitions is still under debate. In a conservative estimate, we assume that bubble acceleration stops long before the bubbles collide.  So we consider the case of fast detonations. Let us stress that in this case almost  the entire released latent heat will go into the fluid, either into heating the plasma or setting it into motion. The scalar field itself is then completely irrelevant when it comes to the production of gravitational waves, but the envelope approximation might still apply.

Assuming that the dynamics of the fluid can be modelled as that of a scalar field in the envelope approximation leads to a gravitational wave signal with maximum amplitude and
peak frequency as \cite{Huber:2008hg, eLISAreport} 
\bea
 h^2 \tilde \Omega_{env}= 1.67\cdot 10^{-5} \tilde \Delta \kappa^2 \left(\frac H \beta\right)^2    \left(\frac {\alpha}{\alpha+1}\right)^2 \left( \frac {g_*}{100}\right)^{-1/3} \ , \\
 \tilde f_{env} = 16.5 \, \mu\textrm{Hz}   \left(\frac {f_{env}}{\beta}\right) \left(\frac \beta H \right) \left(\frac {T_n} {100 \textrm{ GeV}}\right) \left( \frac {g_*}{100}\right)^{1/6} \  ,
\eea
where the parameters $\tilde \Delta, \kappa$ and ${f_{env}}/{\beta}$ are approximated as
\begin{align}
 \kappa&= \frac{\alpha}{0.73 + 0.083 \sqrt{\alpha} + \alpha}     ~,  \\
 \tilde \Delta &= \frac{0.11 v_w^3}{0.42 +v_w^2} \simeq 0.0774 ~,\\
 \frac {f_{env}}{\beta} &= \frac{0.62}{1.8 -0.1 v_w + v_w^2}\simeq 0.230 \ .
\end{align}
Consequently, the spectrum of the GW stochastic background has the
shape given by~\cite{Huber:2008hg}
\begin{align}
 \Omega_{env} (f) = \tilde \Omega_{env} 
\frac{3.8  (f / \tilde f_{env})^{2.8}}
{2.8  + (f/\tilde f_{env})^{3.8}} ~.
\end{align}
Beyond the peak frequency the signal falls as $f^{-1}$.
The GW spectra corresponding to the four benchmark scenarios are
reported in figure~\ref{fig:GWs1}.

\subsection{Gravitational wave signal from sound waves}

Numerical simulations \cite{Hindmarsh:2013xza, Hindmarsh:2015qta} show that the description
of the fluid as a scalar field is too simplistic.\footnote{The scalar field-like description would only apply to the case of strict runaway until bubble collision.} While colliding scalar field bubbles behave very non-linearly, the fluid dynamics turns out to be very linear and can be described 
as an ensemble  of sound waves. These waves are mostly generated when the bubbles collide, but they do not disappear when the phase transition is completed. They are rather expected to be damped mostly by the Hubble expansion. As a result, the gravitational wave amplitude goes
as $(H/\beta)$ rather than $(H/\beta)^2$. This effect should not be viewed as an enhancement
compared to another source.  It is just a more accurate description of one and the same source: the fluid. 

Notice that the contribution from sound waves has been only simulated up to values $\alpha \simeq 0.1$. Besides, it is not certain how long the sound waves persist in the plasma
due to weak shock formation. We consider the envelope approximation and the sound wave approximation as conservative and somewhat optimistic cases of GW production.

The peak amplitude of GW radiation from sound waves is given by~\cite{Hindmarsh:2013xza, Hindmarsh:2015qta, eLISAreport}
\be
 h^2 \tilde \Omega_{sw}= 2.65\cdot 10^{-6} \, v_w \, \kappa^2 \left(\frac H \beta\right)    \left(\frac {\alpha}{\alpha+1}\right)^2 \left( \frac {g_*}{100}\right)^{-1/3} \ , 
\ee
which is larger than the result from the envelope approximation by a factor $\beta/H$. 
The peak frequency is 
\be
 \tilde f_{sw} = 19 \, \mu\textrm{Hz}   \frac{1}{v_w} \left(\frac \beta H \right) \left(\frac {T_n} {100 \textrm{ GeV}}\right) \left( \frac {g_*}{100}\right)^{1/6} \  ,
\ee
and a reasonable fit to the numerical spectrum is given by
\be
\label{Ssw}
\Omega_{sw} (f) = \tilde \Omega_{sw}  \, 
\left(\frac{7}{4 + 3\,(f / \tilde f_{sw})^{2}}  \right)^{7/2}\, 
(f / \tilde f_{sw})^{3}\, .
\ee
Beyond the peak frequency the signal falls off as $f^{-4}$.
The GW spectra corresponding to the four benchmark scenarios are
reported in figure~\ref{fig:GWs2}.

\subsection{Probing the signals at eLISA}

The first GW experiment that might be able to probe the stochastic
background in the mHz range is eLISA. This laser interferometry is
under discussion in the context of the ESA-L3 Gravitational Wave
Mission. The precise experiment architecture is still under debate and
several designs, with their corresponding detection performances, are
under investigation. In the present analysis we consider three
possible experimental scenarios, whose sensitive curves correspond to:
two arms of 2-Gm length and five years of data taking (Design 1); three
arms of 1-Gm length and five years of data taking (Design 2); 
three arms of 5-Gm length and five years of data taking (Design 3).

\begin{figure}
\centering
\includegraphics[width=0.75\textwidth, clip]{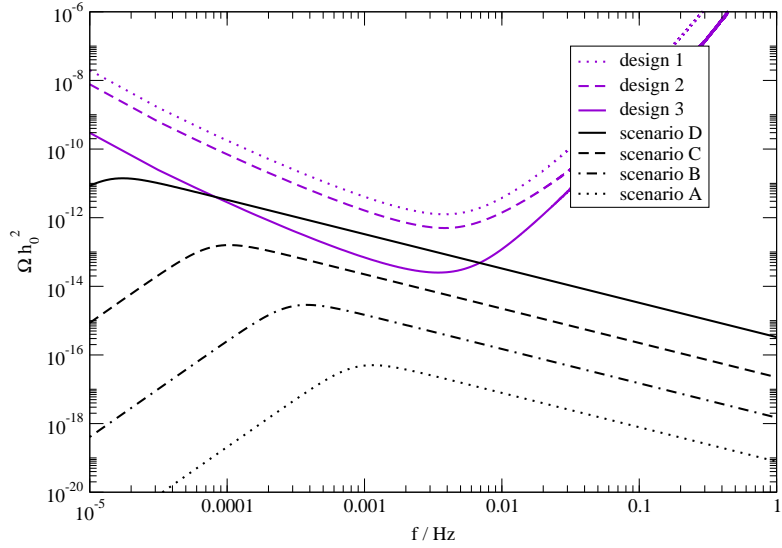}
\caption{Spectrum of the stochastic gravitational wave backgrounds
  (black lines) coming from bubble collisions generated during the
  EWPT in the benchmark points of table \ref{tab:params}. The sensitivity curves of
  the eLISA designs described in the text are displayed in blue.}
\label{fig:GWs1}
\end{figure}

\begin{figure}
\centering
\includegraphics[width=0.75\textwidth, clip]{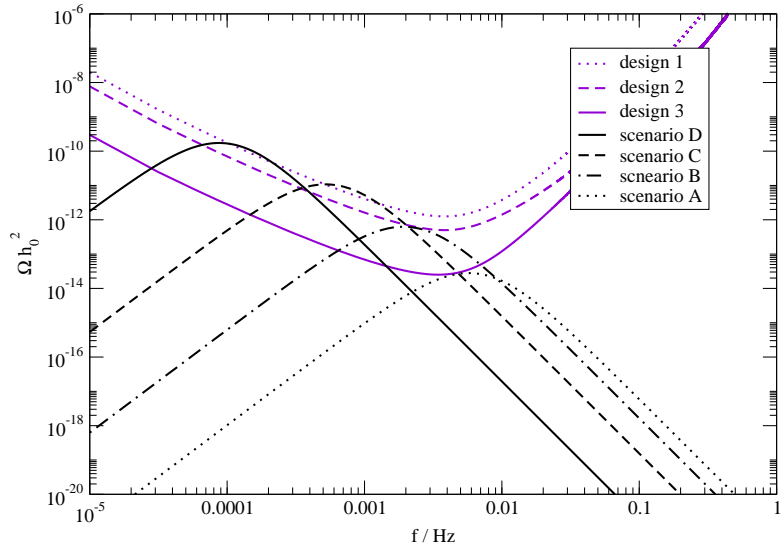}
\caption{As in figure~\ref{fig:GWs1} but for spectrum of the stochastic gravitational wave backgrounds
  (black lines) coming from sound waves.}
\label{fig:GWs2}
\end{figure}

It is important to stress that the sensitivity curves for a stochastic
background are different from the ones calculated for isolated
sources. For the GW signals generated by a first-order phase
transition it is more appropriate to consider sensitivity lines based
on ``power-law integrated curves''~\cite{Thrane:2013oya}. The
power-law sensitivity curves corresponding to the three designs above
have been determined within the eLISA working groups~\cite{internal}
and applied to study the eLISA capabilities to probe the phase transitions~\cite{eLISAreport}. They are
reprinted in figures~\ref{fig:GWs1} and \ref{fig:GWs2} (blue lines)~\footnote{For these
  sensitivity curves the acceleration noise $3\times
  10^{-15}{\rm m \, s}^{-2}{\rm  Hz}^{-1/2}$ is assumed. The signal is assumed to be
  discriminated from the noise as suggested
  in~\cite{Thrane:2013oya,Adams:2010vc,Adams:2013qma}.  We particularly thank Antoine
  Petiteau for calculating the sensitivity curves that we use
  here~\cite{internal}.}.

As figure \ref{fig:GWs1} shows, in the considered benchmark scenarios
the bubble collisions during the EWPT produce sizable GW stochastic
backgrounds. However the detection of the signal depends on the
specific case.  In Scenario D, which is very close to metastability,
the signal is detectable if eLISA is approved with Design 3. Scenario
C is border line with the sensitivity curve of Design 3. Establishing
its detectability by adopting Design 3 is delicate as the conclusion
depends on the uncertainties/assumptions the sensitivity curves are
based on. On the other hand, it seems plausible that improvements in
the data analysis can make Scenario C detectable. Scenario A and B are
instead too weak for eLISA, independently of the experiment design.

Similarly, figure \ref{fig:GWs2} shows that also the sound waves during the 
EWPT produce sizable GW stochastic backgrounds. In this case, even Designs 1 and 2 are able to detect the projected signals in some of the scenarios. The scenarios A, B and C are then within the sensitivity of Design 1, while Scenario D is borderline.

In summary, the GW energy released by the EWPT in our benchmark
scenarios is large. Still in general it seems unlikely that eLISA can probe the EWPT of the 
general NMSSM if the (more economic) option Design 1 is adopted.\\[2mm]
 
As final remark, let us emphasize the differences between the present
analysis and the ones in \cite{NMSSMpreLHC+GW, oldNMSSM} that also
study GW signals in the singlet extension of the MSSM (namely the
nMSSM). First of all, in \cite{oldNMSSM} the parameter scenarios with
a strong EWPT have been identified via a vast scan over the parameter
space (without quadratic and cubic terms in the superpotential) while
here we construct them using an approximate $Z_2$ symmetry as
guideline. This approximated $Z_2$ symmetry, with the singlet having
negligible VEV in the electroweak broken phase, avoids problems with
the LHC Higgs measurements and constitutes a crucial difference from
refs.~\cite{NMSSMpreLHC+GW, oldNMSSM} for what concerns collider
constraints. Moreover the scenarios with the strongest EWPT found in
\cite{oldNMSSM} have somewhat larger $\alpha$ and also larger $\beta$
than those we produce in our benchmark points. We stress that in our
setup we could reach even bigger values of $\alpha$ at the cost of
some tuning but we refrain from doing so. A further difference is that
ref.~\cite{oldNMSSM} is based on two estimates for gravitational waves
caused by turbulence (namely \cite{Dolgov} and
\cite{Durrer}). However, recent numerical
analyses~\cite{Hindmarsh:2013xza, Hindmarsh:2015qta} show that sound
waves dominate the dynamics of the plasma and turbulence is probably
negligible, at least for frequencies around the peak of the GW
spectrum. Last, we use improved sensitivity curves, as published
in~\cite{eLISAreport, internal}, that take into account the stochastic
long-lasting nature of the GW signal and the capabilities of the
possible eLISA configurations. This sensitivity improvement is the
main reason why our conclusion is more optimistic than in
ref.~\cite{oldNMSSM} for what concerns the observation of GWs in the
singlet extension of the MSSM.

\section{Conclusion\label{sec:dis}}

We studied the electroweak phase transition in the general NMSSM. We
found new regions in parameter space with very strong first-order
electroweak phase transitions. In these regions the electroweak
minimum is close to metastability of the symmetric phase. The strong
phase transition occurs in a parameter domain where one of the Higgs
doublets is rather heavy and the scalar potential of the light degrees
of freedom displays an approximate (and accidental) $Z_2$ 
symmetry. The squarks and sleptons were assumed heavy to easily
fulfill all experimental constraints.

The presence of heavy particles leads to large one-loop corrections
that have to be renormalized. In the present analysis, we did not only
impose renormalization conditions in the broken phase of the effective
potential but also in the symmetric one. In this way we ensured that
the approximate $Z_2$ symmetry is still present at the one-loop
level. In general, it guarantees that the properties of the phase
transition are not too strongly affected by the loop contributions.
Although we applied this renormalization method to a specific
supersymmetric model, its application looks promising also for more
generic theories where the EWPT depends radiatively on heavy fields.

We also checked the collider phenomenology of the parameter space
under consideration. With our choice of parameters, the singlet is the
unique new light degree of freedom with a mass around $100$\,GeV. One
pseudoscalar and some of the (Higgsino-like) neutralinos/charginos
have masses of about $300$\,GeV. All the other degrees of freedom
beyond the Standard Model have masses of TeV range. After all, the
arising spectrum is in accord with present collider constraints. 

As anticipated, the model displays a very strong first-order
electroweak phase transition in the identified parameter space. In
part of this parameter region, the phase transition is so strong that
the electroweak broken phase becomes metastable and the system does
not tunnel during the course of the universe. Close to this situation,
we found very strong phase transitions with large Higgs bubbles and
sizable latent heat.

Moreover we analyzed the expected gravitational wave signal. We found
that close to the metastable parameter regime (where the latent heat
is of order of the radiation energy) the gravitational wave signal is
strong. Under some circumstances, the corresponding gravitational wave
stochastic background falls into the sensitivity ballpark of the eLISA
interferometer. However, the detection seems feasible only if the
three-arms eLISA architecture is finally approved.

We stress that the analysis did not aim to be a comprehensive study of
the full available parameter space. The aim was to provide a proof of
principle showing that in supersymmetric models, striking
gravitational wave signals coming from the electroweak phase
transition are possible. For this reason a more complete study of the
parameter space would be worthy. For instance, we imposed an
approximate $Z_2$ but a priori small departures from this
configuration could still lead to sizable gravitational waves. This
would also open the possibility of having the heaviest Higgs below the
TeV scale, provided one of the MSSM-like Higgses is aligned to the SM
one. Moreover, for simplicity we assumed the singlino to be somewhat
heavier than the Higgsinos. By modifying this assumption and allowing
a singlino close in mass to the Higgsinos (which in principle does not
seem in tension with the strong EWPT requirement) would be interesting
in order to tackle the dark matter puzzle. Last but not the least, the
discovered parameter region should have appealing implications for
electroweak baryogenesis. These interesting lines of research are left
for future studies.

\section*{Acknowledgments}

We are grateful to Antoine Petiteau for providing important
information on the eLISA sensitivity curves.

This work was partly supported by the German Science Foundation (DFG) under
the Collaborative Research Center (SFB) 676 Particles, Strings and the Early
Universe.  GN was also partly
supported by the Swiss National Science Foundation (SNF) under grant 200020-155935.

\appendix


\begin{thebibliography}{99}


\bibitem{Kuzmin:1985mm}
  V.~A.~Kuzmin, V.~A.~Rubakov and M.~E.~Shaposhnikov,
  ``On the Anomalous Electroweak Baryon Number Nonconservation in the Early Universe'',
  Phys.\ Lett.\ B {\bf 155} (1985) 36.



\bibitem{witten}
  E.~Witten,
  Phys.\ Rev.\ D {30} (1984) 272;
  C.J.~Hogan,
  Mon.\ Not.\ Roy.\ Astron.\ Soc.\  {218} (1986) 629;
  A.~Kosowsky, M.~S.~Turner and R.~Watkins,
  Phys.\ Rev.\ D {45} (1992) 4514;
  M.~Kamionkowski, A.~Kosowsky and M.S.~Turner,
  Phys.\ Rev.\ D {49} (1994) 2837
  [astro-ph/9310044];


\bibitem{nonpert}
K.~Kajantie, K.~Rummukainen and M.~E.~Shaposhnikov,
  Nucl.\ Phys.\  B {\bf 407}, 356 (1993)
  [arXiv:hep-ph/9305345];
  Z.~Fodor, J.~Hein, K.~Jansen, A.~Jaster and I.~Montvay,
  Nucl.\ Phys.\  B {\bf 439}, 147 (1995)
  [arXiv:hep-lat/9409017];
  K.~Kajantie, M.~Laine, K.~Rummukainen and M.~E.~Shaposhnikov,
  Nucl.\ Phys.\  B {\bf 466}, 189 (1996)
  [arXiv:hep-lat/9510020];
  K.~Jansen,
  Nucl.\ Phys.\ Proc.\ Suppl.\  {\bf 47}, 196 (1996)
  [arXiv:hep-lat/9509018].


\bibitem{Randall:2006py}
  L.~Randall and G.~Servant,
  JHEP {\bf 0705} (2007) 054
  [hep-ph/0607158].
  G.~Nardini, M.~Quiros and A.~Wulzer,
  ``A Confining Strong First-Order Electroweak Phase Transition,''
  JHEP {\bf 0709} (2007) 077
  [arXiv:0706.3388 [hep-ph]];
  T.~Konstandin, G.~Nardini and M.~Quiros,
  ``Gravitational Backreaction Effects on the Holographic Phase Transition,''
  Phys.\ Rev.\ D {\bf 82} (2010) 083513
  [arXiv:1007.1468 [hep-ph]].




\bibitem{mssm1} 
M.~S.~Carena, M.~Quiros and C.~E.~M.~Wagner,
  Phys.\ Lett.\  B {\bf 380} (1996) 81
  [arXiv:hep-ph/9603420];
%
D.~Delepine, J.~M.~Gerard, R.~Gonzalez Felipe and J.~Weyers,
  Phys.\ Lett.\  B {\bf 386} (1996) 183
  [arXiv:hep-ph/9604440];
  Nucl.\ Phys.\  B {\bf 481} (1996) 43
  [Erratum-ibid.\  B {\bf 548} (1999) 673]
  [arXiv:hep-ph/9605283];
 G.~R.~Farrar and M.~Losada,
  Phys.\ Lett.\  B {\bf 406} (1997) 60
  [arXiv:hep-ph/9612346]. 
  D.~Bodeker, P.~John, M.~Laine and M.~G.~Schmidt,
  Nucl.\ Phys.\  B {\bf 497}, 387 (1997)
  [arXiv:hep-ph/9612364];%
 B.~de Carlos and J.~R.~Espinosa,
  Nucl.\ Phys.\  B {\bf 503} (1997) 24
  [arXiv:hep-ph/9703212];
%
  T.~Konstandin, T.~Prokopec, M.~G.~Schmidt and M.~Seco,
  Nucl.\ Phys.\ B {\bf 738} (2006) 1
  [arXiv:hep-ph/0505103].
  V.~Cirigliano, S.~Profumo and M.~J.~Ramsey-Musolf,
  JHEP {\bf 0607} (2006) 002
  [arXiv:hep-ph/0603246];
  A.~De Simone, G.~Nardini, M.~Quiros and A.~Riotto,
  JCAP {\bf 1110} (2011) 030
  [arXiv:1107.4317 [hep-ph]];
 M.~Laine, G.~Nardini and K.~Rummukainen,
  JCAP {\bf 1301} (2013) 011
  [arXiv:1211.7344 [hep-ph]], 
  PoS LATTICE {\bf 2013} (2014) 104
  [arXiv:1311.4424 [hep-lat]].

\bibitem{mssm1bis}
M.~Carena, G.~Nardini, M.~Quiros and C.~E.~M.~Wagner,
  Nucl.\ Phys.\  B {\bf 812} (2009) 243
  [arXiv:0809.3760 [hep-ph]];


\bibitem{mssm2}
  A.~Menon and D.~E.~Morrissey,
  Phys.\ Rev.\ D {\bf 79}, 115020 (2009)
  [arXiv:0903.3038 [hep-ph]].
  T.~Cohen, D.~E.~Morrissey and A.~Pierce,
  arXiv:1203.2924 [hep-ph].
  D.~Curtin, P.~Jaiswal and P.~Meade,
  arXiv:1203.2932 [hep-ph].
 M.~Carena, G.~Nardini, M.~Quiros and C.~E.~M.~Wagner,
  JHEP {\bf 1302} (2013) 001
  [arXiv:1207.6330 [hep-ph]];


\bibitem{Carena:2008rt}
  M.~Carena, G.~Nardini, M.~Quiros and C.~E.~M.~Wagner,
  JHEP {\bf 0810} (2008) 062
  [arXiv:0806.4297 [hep-ph]].



\bibitem{LHCsearchesSUSY}
  D.~P.~Bargassa,
  EPJ Web Conf.\  {\bf 71} (2014) 00010;
 G.~Aad {\it et al.} [ATLAS Collaboration],
  arXiv:1507.05525 [hep-ex];

\bibitem{LHCsearchesEXO}
  A.~Hinzmann [CMS Collaboration],
  EPJ Web Conf.\  {\bf 95} (2015) 04030;
  J.~Tam [ATLAS Collaboration],
  EPJ Web Conf.\  {\bf 95} (2015) 04065.




\bibitem{LHCsingl}
  G.~W.~Anderson and L.~J.~Hall,
  Phys.\ Rev.\ D {\bf 45} (1992) 2685.
  K.~Enqvist, K.~Kainulainen and I.~Vilja,
  Nucl.\ Phys.\ B {\bf 403} (1993) 749;
  S.~W.~Ham, Y.~S.~Jeong and S.~K.~Oh,
  J.\ Phys.\ G {\bf 31} (2005) 857 [hep-ph/0411352]; 
  J.~R.~Espinosa and M.~Quiros,
  Phys.\ Rev.\ D {\bf 76} (2007) 076004
  [hep-ph/0701145];
  A.~Ahriche,
  Phys.\ Rev.\ D {\bf 75} (2007) 083522
  [hep-ph/0701192];
  A.~Ashoorioon and T.~Konstandin,
  JHEP {\bf 0907} (2009) 086
  [arXiv:0904.0353 [hep-ph]];
  J.~R.~Espinosa, T.~Konstandin and F.~Riva,
  Nucl.\ Phys.\ B {\bf 854} (2012) 592
  [arXiv:1107.5441 [hep-ph]].

\bibitem{LHCsingl2}
 V.~Barger, P.~Langacker, M.~McCaskey, M.~J.~Ramsey-Musolf and G.~Shaughnessy,
  Phys.\ Rev.\ D {\bf 77} (2008) 035005
  [arXiv:0706.4311 [hep-ph]];
  J.~M.~No and M.~Ramsey-Musolf,
  Phys.\ Rev.\ D {\bf 89} (2014) 095031
  [arXiv:1310.6035 [hep-ph]];
  S.~Profumo, M.~J.~Ramsey-Musolf, C.~L.~Wainwright and P.~Winslow,
  Phys.\ Rev.\ D {\bf 91} (2015) 3,  035018
  [arXiv:1407.5342 [hep-ph]];
   M.~Jiang, L.~Bian, W.~Huang and J.~Shu,
  arXiv:1502.07574 [hep-ph].

\bibitem{Barger:2006dh}
  V.~Barger, P.~Langacker, H.~S.~Lee and G.~Shaughnessy,
  Phys.\ Rev.\ D {\bf 73} (2006) 115010
  [hep-ph/0603247];
  G.~G.~Ross, K.~Schmidt-Hoberg and F.~Staub,
  JHEP {\bf 1208} (2012) 074
  [arXiv:1205.1509 [hep-ph]].


\bibitem{eLISAreport}
  C.~Caprini {\it et al.},
  arXiv:1512.06239 [astro-ph.CO].



\bibitem{NMSSMpreLHC+GW}
  R.~Apreda, M.~Maggiore, A.~Nicolis and A.~Riotto,
  Nucl.\ Phys.\ B {\bf 631} (2002) 342
  [gr-qc/0107033].


\bibitem{oldNMSSM}
S.~J.~Huber and T.~Konstandin,
  JCAP {\bf 0805} (2008) 017
  [arXiv:0709.2091 [hep-ph]].
 

\bibitem{NMSSMpreLHC}
  A.~Menon, D.~E.~Morrissey and C.~E.~M.~Wagner,
  Phys.\ Rev.\ D {\bf 70} (2004) 035005
  [hep-ph/0404184];
C.~Balazs, M.~Carena, A.~Freitas and C.~E.~M.~Wagner,
  JHEP {\bf 0706} (2007) 066
  [arXiv:0705.0431 [hep-ph]];
  W.~Huang, Z.~Kang, J.~Shu, P.~Wu and J.~M.~Yang,
  Phys.\ Rev.\ D {\bf 91} (2015) 2,  025006
  [arXiv:1405.1152 [hep-ph]];
   M.~Kakizaki, S.~Kanemura and T.~Matsui,
  Phys.\ Rev.\ D {\bf 92} (2015) 11,  115007
  doi:10.1103/PhysRevD.92.115007
  [arXiv:1509.08394 [hep-ph]].


\bibitem{Kozaczuk:2014kva}
  J.~Kozaczuk, S.~Profumo, L.~S.~Haskins and C.~L.~Wainwright,
  JHEP {\bf 1501} (2015) 144
  [arXiv:1407.4134 [hep-ph]].



\bibitem{Staub:2013tta}
  F.~Staub,
  Comput.\ Phys.\ Commun.\  {\bf 185} (2014) 1773
  [arXiv:1309.7223 [hep-ph]].



\bibitem{Masina:2015ixa}
  I.~Masina, G.~Nardini and M.~Quiros,
  Phys.\ Rev.\ D {\bf 92} (2015) 3,  035003
  [arXiv:1502.06525 [hep-ph]].

\bibitem{Cornwall:1974vz}
  J.~M.~Cornwall, R.~Jackiw and E.~Tomboulis,
  Phys.\ Rev.\ D {\bf 10} (1974) 2428.





\bibitem{Degrassi:2009yq}
  G.~Degrassi and P.~Slavich,
  Nucl.\ Phys.\ B {\bf 825} (2010) 119
  [arXiv:0907.4682 [hep-ph]].

\bibitem{Staub:2010ty}
  F.~Staub, W.~Porod and B.~Herrmann,
  JHEP {\bf 1010} (2010) 040
  [arXiv:1007.4049 [hep-ph]].

\bibitem{Ender:2011qh}
  K.~Ender, T.~Graf, M.~Muhlleitner and H.~Rzehak,
  Phys.\ Rev.\ D {\bf 85} (2012) 075024
  [arXiv:1111.4952 [hep-ph]].

\bibitem{Graf:2012hh}
  T.~Graf, R.~Grober, M.~Muhlleitner, H.~Rzehak and K.~Walz,
  JHEP {\bf 1210} (2012) 122
  [arXiv:1206.6806 [hep-ph]].

\bibitem{Staub:2015aea}
  F.~Staub, P.~Athron, U.~Ellwanger, R.~Grober, M.~Muhlleitner, P.~Slavich and A.~Voigt,
  arXiv:1507.05093 [hep-ph].




  \bibitem{TuningNMSSM}
  U.~Ellwanger, G.~Espitalier-Noel and C.~Hugonie,
  JHEP {\bf 1109} (2011) 105
  [arXiv:1107.2472 [hep-ph]];
  Nucl.\ Phys.\ B {\bf 862} (2012) 710
  [arXiv:1108.1284 [hep-ph]];
  K.~Agashe, Y.~Cui and R.~Franceschini,
  JHEP {\bf 1302} (2013) 031
  [arXiv:1209.2115 [hep-ph]].
 






\bibitem{Aad:2015zhl}
  G.~Aad {\it et al.} [ATLAS and CMS Collaborations],
  Phys.\ Rev.\ Lett.\  {\bf 114} (2015) 191803
  [arXiv:1503.07589 [hep-ex]].


\bibitem{Aad:2015gba}
  G.~Aad {\it et al.} [ATLAS Collaboration],
  arXiv:1507.04548 [hep-ex];
 CMS Collaboration [CMS Collaboration],
and studies of the compatibility of its couplings with the standard model,''
  CMS-PAS-HIG-14-009.



\bibitem{Delgado:2013zfa}
   J.~F.~Gunion and H.~E.~Haber,
  Phys.\ Rev.\ D {\bf 67} (2003) 075019
  [hep-ph/0207010];
  A.~Delgado, G.~Nardini and M.~Quiros,
  JHEP {\bf 1307} (2013) 054
  [arXiv:1303.0800 [hep-ph]];
   M.~Carena, I.~Low, N.~R.~Shah and C.~E.~M.~Wagner,
  JHEP {\bf 1404} (2014) 015
  [arXiv:1310.2248 [hep-ph]];
P.~S.~B.~Dev and A.~Pilaftsis,
  JHEP {\bf 1412} (2014) 024
  [arXiv:1408.3405 [hep-ph]].



\bibitem{Giardino:2013bma}
  P.~P.~Giardino, K.~Kannike, I.~Masina, M.~Raidal and A.~Strumia,
  JHEP {\bf 1405} (2014) 046
  [arXiv:1303.3570 [hep-ph]];
 A.~Falkowski, F.~Riva and A.~Urbano,
  JHEP {\bf 1311} (2013) 111
  [arXiv:1303.1812 [hep-ph]].


\bibitem{Bomark}
    N.~E.~Bomark, S.~Moretti, S.~Munir and L.~Roszkowski,
  JHEP {\bf 1502} (2015) 044
  [arXiv:1409.8393 [hep-ph]].

\bibitem{Tesi}
  D.~Buttazzo, F.~Sala and A.~Tesi,
  arXiv:1505.05488 [hep-ph].

\bibitem{Tania}
 T.~Robens and T.~Stefaniak,
  Eur.\ Phys.\ J.\ C {\bf 75} (2015) 104
  [arXiv:1501.02234 [hep-ph]].




\bibitem{Berggren:2013vfa}
  M.~Berggren, F.~Brümmer, J.~List, G.~Moortgat-Pick, T.~Robens, K.~Rolbiecki and H.~Sert,
  Eur.\ Phys.\ J.\ C {\bf 73} (2013) 12,  2660
  [arXiv:1307.3566 [hep-ph]].



\bibitem{Enberg:2015qwa}
  R.~Enberg, S.~Munir, C.~Pérez de los Heros and D.~Werder,
  arXiv:1506.05714 [hep-ph].

\bibitem{Kim:2014noa}
  J.~S.~Kim and T.~S.~Ray,
  Eur.\ Phys.\ J.\ C {\bf 75} (2015) 40
  [arXiv:1405.3700 [hep-ph]].



\bibitem{Ellwanger:2009dp}
  U.~Ellwanger, C.~Hugonie and A.~M.~Teixeira,
  Phys.\ Rept.\  {\bf 496} (2010) 1
  [arXiv:0910.1785 [hep-ph]].




\bibitem{Coleman:1977py}
  S.~R.~Coleman,
  Phys.\ Rev.\ D {\bf 15} (1977) 2929
   [Phys.\ Rev.\ D {\bf 16} (1977) 1248].

\bibitem{Callan:1977pt}
  C.~G.~Callan, Jr. and S.~R.~Coleman,
  Phys.\ Rev.\ D {\bf 16} (1977) 1762.

\bibitem{Linde:1980tt}
  A.~D.~Linde,
  Phys.\ Lett.\ B {\bf 100} (1981) 37.






\bibitem{Huber:2008hg}
  S.~J.~Huber and T.~Konstandin,
  JCAP {\bf 0809} (2008) 022
  [arXiv:0806.1828 [hep-ph]].



\bibitem{Hindmarsh:2013xza}
  M.~Hindmarsh, S.~J.~Huber, K.~Rummukainen and D.~J.~Weir,
  Phys.\ Rev.\ Lett.\  {\bf 112} (2014) 041301
  [arXiv:1304.2433 [hep-ph]].


\bibitem{Hindmarsh:2015qta}
  M.~Hindmarsh, S.~J.~Huber, K.~Rummukainen and D.~J.~Weir,
  arXiv:1504.03291 [astro-ph.CO].





\bibitem{MoAndPro}
 G.~D.~Moore and T.~Prokopec,
  ``How fast can the wall move? A Study of the 
  electroweak phase transition dynamics,''
  Phys.\ Rev.\ D\ {\bf 52} (1995) 7182
  [hep-ph/9506475].

\bibitem{John:2000zq}
  P.~John and M.~G.~Schmidt,
  Nucl.\ Phys.\ B {\bf 598} (2001) 291
   [Erratum-ibid.\ B {\bf 648} (2003) 449]
  [hep-ph/0002050].

\bibitem{Megevand:2009gh}
  A.~Megevand and A.~D.~Sanchez,
  Nucl.\ Phys.\ B {\bf 825} (2010) 151
  [arXiv:0908.3663 [hep-ph]].

\bibitem{energybudget}
  J.~R.~Espinosa, T.~Konstandin, J.~M.~No and G.~Servant,
  ``Energy Budget of Cosmological First-order Phase Transitions,''
  JCAP {\bf 1006} (2010) 028
  [arXiv:1004.4187 [hep-ph]].


\bibitem{Megevand:2012rt}
  A.~Megevand and A.~D.~Sanchez,
  Nucl.\ Phys.\ B {\bf 865} (2012) 217
  [arXiv:1206.2339 [astro-ph.CO]].

\bibitem{Huber:2013kj}
  S.~J.~Huber and M.~Sopena,
  arXiv:1302.1044 [hep-ph].


\bibitem{Konstandin:2014zta}
  T.~Konstandin, G.~Nardini and I.~Rues,
  JCAP {\bf 1409} (2014) 09,  028
  [arXiv:1407.3132 [hep-ph]];

\bibitem{Kozaczuk:2015owa}
  J.~Kozaczuk,
  JHEP {\bf 1510} (2015) 135
  [arXiv:1506.04741 [hep-ph]].



  
\bibitem{Bodeker:2009qy}
  D.~Bodeker and G.~D.~Moore,
  JCAP {\bf 0905} (2009) 009
  [arXiv:0903.4099 [hep-ph]].



\bibitem{Thrane:2013oya}
  E.~Thrane and J.~D.~Romano,
  Phys.\ Rev.\ D {\bf 88} (2013) 12,  124032
  [arXiv:1310.5300 [astro-ph.IM]].


\bibitem{internal}
  A.~Petiteau,
  to appear.



\bibitem{Adams:2010vc}
  M.~R.~Adams and N.~J.~Cornish,
  Phys.\ Rev.\ D {\bf 82} (2010) 022002
  [arXiv:1002.1291 [gr-qc]].

\bibitem{Adams:2013qma}
  M.~R.~Adams and N.~J.~Cornish,
  Phys.\ Rev.\ D {\bf 89} (2014) 2,  022001
  [arXiv:1307.4116 [gr-qc]].


 \bibitem{Dolgov}
A.~D.~Dolgov, D.~Grasso and A.~Nicolis,
  Phys.\ Rev.\ D {\bf 66} (2002) 103505
  [astro-ph/0206461].

 \bibitem{Durrer}
 C.~Caprini and R.~Durrer,
  Phys.\ Rev.\ D {\bf 74} (2006) 063521
  [astro-ph/0603476].



\end{thebibliography}
\end{document}